\documentclass[twocolumn,floatfix,showpacs,superscriptaddress]{revtex4}
\usepackage{graphicx}
\usepackage{amsmath}
\usepackage{amssymb}
\usepackage{bm}

\begin{document}

\title{Effect of chiral symmetry on chaotic scattering from Majorana zero modes}
\author{H. Schomerus}
\affiliation{Department of Physics, Lancaster University, LA1 4YB Lancaster, United Kingdom}
\affiliation{Instituut-Lorentz, Universiteit Leiden, P.O. Box 9506, 2300 RA Leiden, The Netherlands}
\author{M. Marciani}
\affiliation{Instituut-Lorentz, Universiteit Leiden, P.O. Box 9506, 2300 RA Leiden, The Netherlands}
\author{C. W. J. Beenakker}
\affiliation{Instituut-Lorentz, Universiteit Leiden, P.O. Box 9506, 2300 RA Leiden, The Netherlands}
\date{April 2015}

\begin{abstract}
In many of the experimental systems that may host Majorana zero modes, a so-called chiral symmetry exists that protects overlapping zero modes from splitting up. This symmetry is operative in a superconducting nanowire that is narrower than the spin-orbit scattering length, and at the Dirac point of a superconductor/topological insulator heterostructure. Here we show that chiral symmetry strongly modifies the dynamical and spectral properties of a chaotic scatterer, even if it binds only a single zero mode. These properties are quantified by the Wigner-Smith time-delay matrix $Q=-i\hbar S^\dagger dS/dE$, the Hermitian energy derivative of the scattering matrix, related to the density of states by $\rho=(2\pi\hbar)^{-1}\,{\rm Tr}\,Q$. We compute the probability distribution of $Q$ and $\rho$, dependent on the number $\nu$ of Majorana zero modes, in the chiral ensembles of random-matrix theory. Chiral symmetry is essential for a significant $\nu$-dependence.
\end{abstract}

\maketitle

In classical mechanics the duration $\tau$ of a scattering process can be defined without ambiguity, for example as the energy derivative of the action. The absence of a quantum mechanical operator of time complicates the simple question ``by how much is an electron delayed?'' \cite{But02,Car02}. Since the action, in units of $\hbar$, corresponds to the quantum mechanical phase shift $\phi$, the quantum analogue of the classical definition is $\tau=\hbar d\phi/dE$. In a multi-channel scattering process, described by an $N\times N$ unitary scattering matrix $S(E)$, one then has a set of delay times $\tau_1,\tau_2,\ldots\tau_N$, defined as the eigenvalues of the so-called Wigner-Smith matrix
\begin{equation}
Q=-i\hbar S^{\dagger}(dS/dE).\label{Qdef}
\end{equation}
(For a scalar $S=e^{i\phi}$ the single-channel definition is recovered.)

This dynamical characterization of quantum scattering processes goes back to work by Wigner and others \cite{Eis48,Wig55,Smi60} in the 1950's. Developments in the random-matrix theory of chaotic scattering from the 1990's \cite{Blu90,Smi91} allowed for a universal description of the statistics of the delay times $\tau_n$ in an ensemble of chaotic scatterers. The inverse delay matrix $Q^{-1}$ turns out to be statistically equivalent to a so-called Wishart matrix \cite{For10}: the Hermitian positive-definite matrix product $WW^\dagger$, with $W$ a rectangular matrix having independent Gaussian matrix elements. The corresponding probability distribution of the inverse delay times $\gamma_n\equiv 1/\tau_n>0$ (measured in units of the Heisenberg time $\tau_{\rm H}=2\pi\hbar/\delta_0$, with mean level spacing $\delta_0$), takes the form \cite{Bro97,note1}
\begin{equation}
P(\{\gamma_n\})\propto\prod_{j>i=1}^N|\gamma_i-\gamma_j|^\beta\prod_{k=1}^N \gamma_k^{\beta N/2}e^{-\beta\tau_{\rm H}\gamma_k/2}.\label{Pgammadef}
\end{equation}
The symmetry index $\beta\in\{1,2,4\}$ distinguishes real, complex, and quaternion Hamiltonians. This connection between delay-time statistics and the Wishart ensemble is the dynamical counterpart of the connection between spectral statistics and the Wigner-Dyson ensemble \cite{Wig67,Dys62} --- discovered several decades later although the Wishart ensemble \cite{Wis28} is several decades older than the Wigner-Dyson ensemble.

The delay-time distribution \eqref{Pgammadef} assumes ballistic coupling of the $N$ scattering channels to the outside world. It has been generalized to coupling via a tunnel barrier \cite{Som01,Sav01}, and has been applied to a variety of transport properties (such as thermopower, low-frequency admittance, charge relaxation resistance) of disordered electronic quantum dots and chaotic microwave cavities \cite{Gop96,Bro97b,God99,Kot02,Cra02,Sav03,But05,Nig08,Tex13,Abb13,Mez13,Kui14,Gra14,Cun14}. Because the density of states $\rho(E)$ is directly related to the Wigner-Smith matrix,
\begin{equation}
\rho(E)=(2\pi\hbar)^{-1}\,{\rm Tr}\,Q(E)=\textstyle{\sum_{n}}(2\pi\hbar\gamma_n)^{-1},\label{rhoQ}
\end{equation}
the delay-time distribution also provides information on the degree to which levels are broadened by coupling to a continuum.

The discovery of topological insulators and superconductors \cite{Has10,Qi11} has opened up a new arena of applications of random-matrix theory \cite{Ryu10,Bee14}. Topologically nontrivial chaotic scatterers are distinguished by a topological invariant $\nu$ that is either a parity index, $\nu\in \mathbb{Z}_2$, or a winding number, $\nu\in\mathbb{Z}$. In the spectral statistics, topologically distinct systems are immediately identified through the number of zero modes, a total of $|\nu|$ levels pinned to the middle of the excitation gap \cite{Boc00,Iva02}. If the gap is induced by a superconductor, the zero modes are Majorana, of equal electron and hole character \cite{Ali12,Bee13,Ell14}.

These developments raise the question how topological invariants connect to the Wishart ensemble: How do Majorana zero modes affect the dynamics of chaotic scattering? That is the problem adressed and solved in this paper, building on two earlier works \cite{Fyo04,Mar14}. In Ref.\ \onlinecite{Mar14} it was found that a $\mathbb{Z}_2$ invariant (only particle-hole symmetry, symmetry class D in the Altland-Zirnbauer classification \cite{Alt97}) has no effect on the delay-time distribution for ideal (ballistic) coupling to the scatterer: The distribution is the same with or without an unpaired Majorana zero mode in the spectrum. Here we show that the $\mathbb{Z}$ invariant of $|\nu|$-fold degenerate Majorana zero modes does significantly affect the delay-time distribution. This is symmetry class BDI, with particle-hole symmetry as well as chiral symmetry \cite{Ver93,Ver00}. Chiral symmetry without particle-hole symmetry, symmetry class AIII, was considered in Ref.\ \onlinecite{Fyo04} for a scalar $S=e^{i\phi}$, with a single delay time $\tau=\hbar d\phi/dE$. While our interest here is in Majorana modes, for which particle-hole symmetry is essential, our general results include a multi-channel generalization of Ref.\ \onlinecite{Fyo04} and also cover class CII (see the Appendix for additional details).

Majorana zero modes are being pursued in either two-dimensional (2D) or one-dimensional (1D) systems \cite{Ali12,Bee13,Lei12,Sta13}. In the former geometry the zero modes are bound to a magnetic vortex core, in the latter geometry they appear at the end point of a nanowire. Particle-hole symmetry by itself can only protect a single zero mode, so even though the Majoranas always come in pairs, they have to be widely separated. The significance of chiral symmetry is that it provides additional protection for multiple overlapping Majorana zero modes \cite{Fid10,Tur11,Man12,Dga14}. The origin of the chiral symmetry is different in the 1D and 2D geometries.

By definition, chiral symmetry means that the Hamiltonian $H$ anticommutes with a unitary operator. The 1D realization of chiral symmetry relies on the fact that the Rashba Hamiltonian of a nanowire in a parallel magnetic field is real --- if its width $W$ is well below the spin-orbit scattering length. Particle-hole symmetry $H=-\tau_x H^\ast \tau_x$ then implies that $H$ anticommutes with the Pauli matrix $\tau_x$ that switches electrons and holes. It follows that a nanowire with $W\lesssim l_{\rm so}$ (the typical regime of operation) is in the BDI symmetry class and supports multiple degenerate Majorana zero modes at its end \cite{Tew12,Die12,Hui14}.

\begin{figure}[tb]
\centerline{\includegraphics[width=0.9\linewidth]{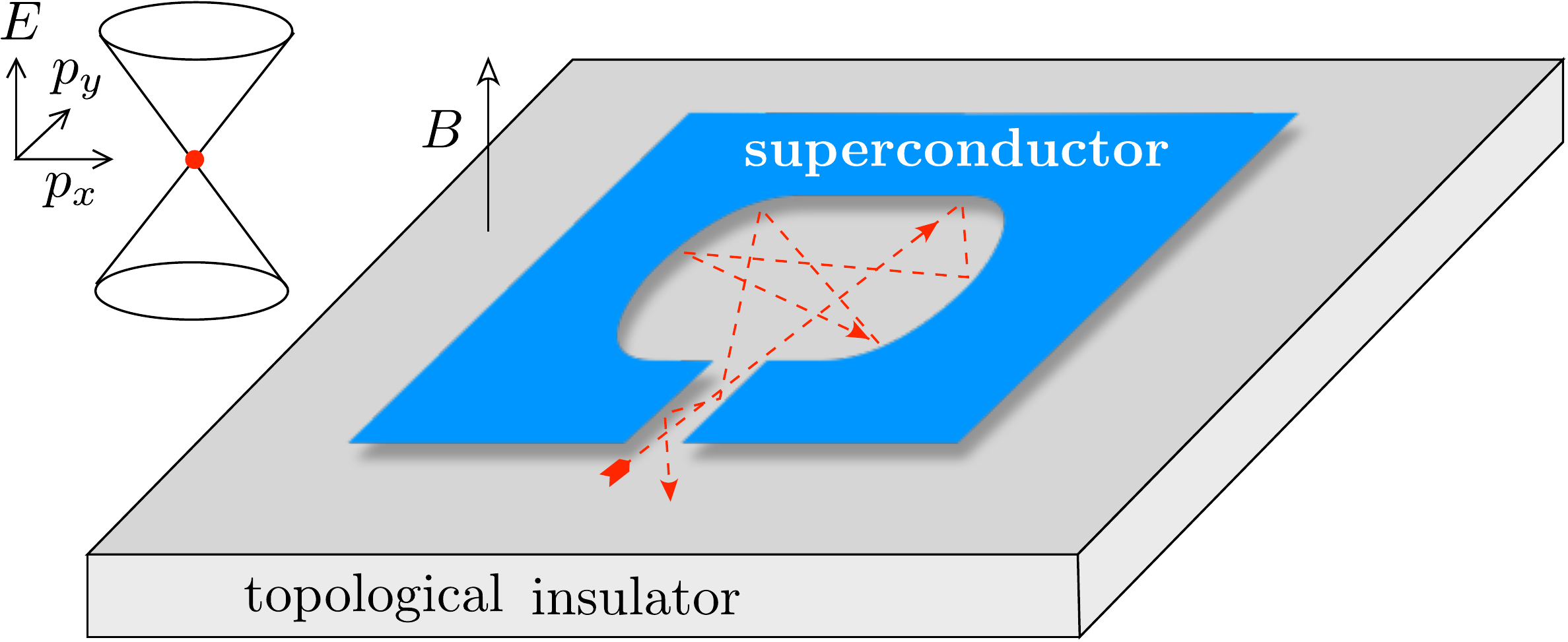}}
\caption{Andreev billiard on the conducting surface of a three-dimensional topological insulator in a magnetic field. The winding number $\nu$ of the superconducting order parameter around the billiard is associated with $|\nu|$ Majorana zero modes, that affect the quantum delay time when the Fermi level lines up with the Dirac point (red dot) of the conical band structure.
}
\label{fig_layout}
\end{figure}

The Andreev billiard of Fig.\ \ref{fig_layout} illustrates a 2D realization on the surface of a topological insulator. The massless Dirac fermions on the surface have a chiral symmetry at the charge-neutrality point (the Dirac point), because the 2D Dirac Hamiltonian
\begin{equation}
H_0=v(p_x-eA_x)\sigma_x+v(p_y-eA_y)\sigma_y \label{DiracH0}
\end{equation}
anticommutes with the Pauli spin-matrix $\sigma_z$. The coupling to a superconducting pair potential $\Delta$ introduces particle-hole symmetry without breaking the chiral symmetry, since the Bogoliubov-De Gennes Hamiltonian
\begin{equation}
H=\begin{pmatrix}
H_0-\mu&-i\sigma_y\Delta\\
i\sigma_y\Delta^\ast &\mu-H_0^\ast
\end{pmatrix}\label{HBDG}
\end{equation}
still anticommutes with $\sigma_z$ for $\mu=0$.

Therefore, overlapping Majorana zero modes in a superconductor/topological insulator heterostructure (the Fu-Kane model \cite{Fu08}) will not split when the chemical potential is tuned to within a Thouless energy $N\delta_0$ from the Dirac point \cite{Che10,Teo10,Cio14}. In this 2D geometry one needs random scattering by disorder to produce a finite density of states at $E=0$, but in order to preserve the chiral symmetry the disorder cannot be electrostatic ($V$ must remain zero). Scattering by a random vector potential is one possibility \cite{Lud94,Mot02}, or alternatively scattering by random surface deformations \cite{Lee09,Dah10,Par11}. To be definite, we will refer to the 2D Andreev billiard geometry in the following, but our results apply as well to 1D nanowires \cite{note0}.

The unitary scattering matrix $S(E)$ of the Andreev billiard is obtained from the Green's function ${\cal G}(E)=K(E-H)^{-1}K^\dagger$ via
\begin{equation}
S(E)=[1-i\pi {\cal G}(E)][1+i\pi{\cal G}(E)]^{-1}.\label{SGErelation}
\end{equation}
The matrix $K$ describes the coupling of the quasibound states inside the billiard to the continuum outside via $2N$ scattering channels \cite{spincounting}. We assume that $K$ commutes with $\sigma_z$ so as not to spoil the chiral symmetry of the Green's function and scattering matrix,
\begin{equation}
\sigma_z{\cal G}(E)=-{\cal G}(-E)\sigma_z\Rightarrow \sigma_z S(E)=S^\dagger(-E)\sigma_z.\label{GESchiral}
\end{equation}
It follows that the matrix product $S_0=\sigma_z S(0)$ is both Hermitian and unitary, so its eigenvalues can only be $+1$ or $-1$. There are $N_\pm=N\pm\nu_0$ eigenvalues equal to $\pm 1$, where the so-called matrix signature $\nu_0$ is determined by the number of Majorana zero modes \cite{Ful11}:
\begin{equation}
\nu_0=\tfrac{1}{2}\,{\rm Tr}\,S_0=\begin{cases}
\nu&{\rm if}\;|\nu|\leq N,\\
N\,({\rm sign}\,\nu)&{\rm if}\;|\nu|\geq N.
\end{cases}.\label{nudef}
\end{equation}

At the Fermi level, the time-delay matrix \eqref{Qdef} depends on $S_0$ and on the first-order energy variation, $\sigma_z S(E)=S_0\cdot[1+iES_1+{\cal O}(E^2)]$. Unitarity requires that $S_1$ is Hermitian and the chiral symmetry \eqref{GESchiral} then implies that $S_1$ commutes with $S_0$. Since $Q(0)\equiv Q_0=\hbar S_1$, the same applies to the time-delay matrix at the Fermi level: $S_0 Q_0=Q_0 S_0$. This implies the block structure
\begin{equation}
S_0=U_0\begin{pmatrix}
\openone_{N_+}&0\\
0&-\openone_{N_-}
\end{pmatrix}U_0^\dagger,\;\;Q_0=U_0\begin{pmatrix}
Q_+&0\\
0&Q_-
\end{pmatrix}U_0^\dagger,\label{S0Q0}
\end{equation}
with $\openone_n$ the $n\times n$ unit matrix, $U_0$ a $2N\times 2N$ unitary matrix, and $Q_\pm$ a pair of $N_\pm\times N_\pm$ Hermitian matrices. There are therefore two sets of delay times $\tau_n^\pm$, $n=1,2,\ldots N_\pm$, corresponding to an eigenvalue $\pm 1$ of $S_0$.

After these preparations we can now state our central result: For ballistic coupling the two matrices $Q_+^{-1}$ and $Q_-^{-1}$ are statistically independent, each described by its own Wishart ensemble \cite{note5} and eigenvalue distribution $P_\pm$ of $\gamma_n^\pm=1/\tau^\pm_n$ given by
\begin{align}
P_\pm(\{\gamma_n^\pm\})\propto{}&\prod_{j>i=1}^{N_\pm}|\gamma^\pm_i-\gamma^\pm_j|^\beta \prod_{k=1}^{N_\pm} (\gamma_k^\pm)^{\beta/2-1}e^{-\beta\tau_{\rm H}\gamma^\pm_k/4}\nonumber\\
&\qquad\qquad\times (\gamma_k^\pm)^{(\beta/2)|\pm\nu-N|},\label{Pgammapmdef}
\end{align}
with symmetry index $\beta=1$ for the class BDI Hamiltonian \eqref{HBDG}. The distribution \eqref{Pgammapmdef} holds also for $|\nu|\geq N$, when the scattering matrix signature \eqref{nudef} is saturated. In that case a single Wishart ensemble remains for all $2N$ delay times, with distribution
\begin{align}
P(\{\gamma_n\})\propto{}&\prod_{j>i=1}^{2N}|\gamma_i-\gamma_j|^\beta \prod_{k=1}^{2N} \gamma_k^{\beta/2-1}e^{-\beta\tau_{\rm H}\gamma_k/4}\nonumber\\
&\qquad\qquad\times \gamma_k^{(\beta/2)(|\nu|-N)},\;\;|\nu|\geq N.\label{Pgammadefsaturated}
\end{align}

The derivation of Eq.\ \eqref{Pgammapmdef} starts from the Gaussian ensemble for Hamiltonians with chiral symmetry \cite{For10,Ver00},
\begin{equation}
H=\begin{pmatrix}
0&{\cal A}\\
{\cal A}^\dagger&0
\end{pmatrix},\;\;
P({\cal A})\propto\exp\left(-\frac{\beta\pi^2}{8\delta_0^{2}{\cal N}}\,{\rm Tr}\,{\cal A}{\cal A}^\dagger\right).
\label{calHdef}
\end{equation}
The rectangular matrix ${\cal A}$ has dimensions ${\cal N}\times({\cal N}+\nu)$, so $H$ has $|\nu|$ eigenvalues pinned to zero. The matrix elements of ${\cal A}$ are real ($\beta=1$, symmetry class BDI, chiral orthogonal ensemble), complex ($\beta=2$, class AIII, chiral unitary ensemble) or quaternion ($\beta=4$, class CII, chiral symplectic ensemble).

The coupling matrix $K=K_1\oplus K_2$ is composed of two rectangular blocks of dimensions $N\times{\cal N}$ and $N\times({\cal N}+\nu)$, having nonzero matrix elements
\begin{equation}
(K_{1})_{nn}=(K_{2})_{nn}=\kappa_n,\;\;n=1,2,\ldots N,\label{Wnndef}
\end{equation}
with $\kappa_n=\sqrt{2{\cal N}\delta_0/\pi^2}\equiv\kappa_0$ for ballistic coupling. These matrices determine the time-delay matrix \eqref{Qdef} via Eq.\ \eqref{SGErelation}. At the Fermi level one has
\begin{equation}
Q_0=2\pi\hbar\Omega\Omega^\dagger,\;\;\Omega=K(H+i\pi K^\dagger K)^{-1}.\label{Q0Omega}
\end{equation}
We seek the distribution of $Q_0$ given the Gaussian distribution of $H$, in the limit ${\cal N}\rightarrow\infty$ at fixed $\nu$.

The corresponding problem in the absence of chiral symmetry was solved \cite{Bro97,Mar14} by using the unitary invariance of the distribution to perform the calculation in the limit $S\rightarrow -1$, when a major simplification occurs. Here this would only work in the topologically trivial case $\nu_0=0$ \cite{note2}, so a different approach is needed. We would like to exploit the block decomposition \eqref{calHdef} of the Hamiltonian, but this decomposition is lost in Eq.\ \eqref{Q0Omega}.

Unitary invariance does allow us to directly obtain the distribution of the eigenvectors of $Q_{\pm}=U_\pm\,{\rm diag}\,(\tau_1^\pm ,\tau_2^\pm,\ldots)U_\pm^\dagger$. From the invariance $P(S_0,Q_0)=P(VS_0 V^\dagger,VQ_0 V^\dagger)$ under joint unitary transformations of $S_0$ and $Q_0$ we conclude that the matrices of eigenvectors $U_0,U_+,U_-$ are all independent and uniformly distributed in the unitary group for $\beta=2$, and in the orthogonal or symplectic subgroups for $\beta=1$ or $\beta=4$.

The ``trick'' that allows us to obtain the eigenvalue distribution is to note that $\tilde{Q}_0=2\pi\hbar\Omega^\dagger\Omega$ has the same nonzero eigenvalues as $Q_0$ --- but unlike $Q_0$ it is block-diagonal:
\begin{subequations}
\label{QLambda}
\begin{align}
&\tilde{Q}_0=2\pi\hbar \begin{pmatrix}
\Lambda_-^{-1}&0\\
0&\Lambda_+^{-1}
\end{pmatrix},\label{QLambdaa}\\
&\Lambda_-= \pi^2 K_1^\dagger K_1+{\cal A}(K_2^\dagger K_2+\epsilon)^{-1}{\cal A}^\dagger,\label{QLambdab}\\
&\Lambda_+= \pi^2 K_2^\dagger K_2+{\cal A}^\dagger(K_1^\dagger K_1+\epsilon)^{-1}{\cal A}.\label{QLambdac}
\end{align}
\end{subequations}
The infinitesimal $\epsilon$ is introduced to regularize the inversion of the singular matrices $K_n^\dagger K_n=\kappa_0^2{\cal P}_n$, where $({\cal P}_{n})_{ij}=1$ if $1\leq i=j\leq N$ and zero otherwise. In the limit $\epsilon\rightarrow 0$ some eigenvalues of $\Lambda_\pm$ diverge, while the others converge to the inverse delay times $\gamma_n^\pm$.

The calculation of the eigenvalues of $\Lambda_\pm$ in the $\epsilon\rightarrow 0$ limit is now a matter of perturbation theory (for details see the Appendix). This is a degenerate perturbation expansion in the null space of ${\cal A}(\openone_{{\cal N}+\nu}-{\cal P}_2){\cal A}^\dagger$ for $\Lambda_+$ and in the null space of ${\cal A}^\dagger(\openone_{\cal N}-{\cal P}_1){\cal A}$ for $\Lambda_-$. The small perturbation (an order $\epsilon$ smaller than the leading order term) is $\pi^2 \kappa_0^2{\cal P}_1+\kappa_0^{-2}{\cal A}{\cal P}_2{\cal A}^\dagger$ and $\pi^2 \kappa_0^2{\cal P}_2+\kappa_0^{-2}{\cal A}^\dagger{\cal P}_1{\cal A}$, for $\Lambda_+$ and $\Lambda_-$ respectively. The Gaussian distribution \eqref{calHdef} of the matrix elements of ${\cal A}$ results in the eigenvalue distributions $P(\{\gamma_n\})=P_+(\{\gamma_n^+\})P_-(\{\gamma_n^-\})$ given by Eq.\ \eqref{Pgammapmdef}.

\begin{figure}[tb]
\centerline{\includegraphics[width=1\linewidth]{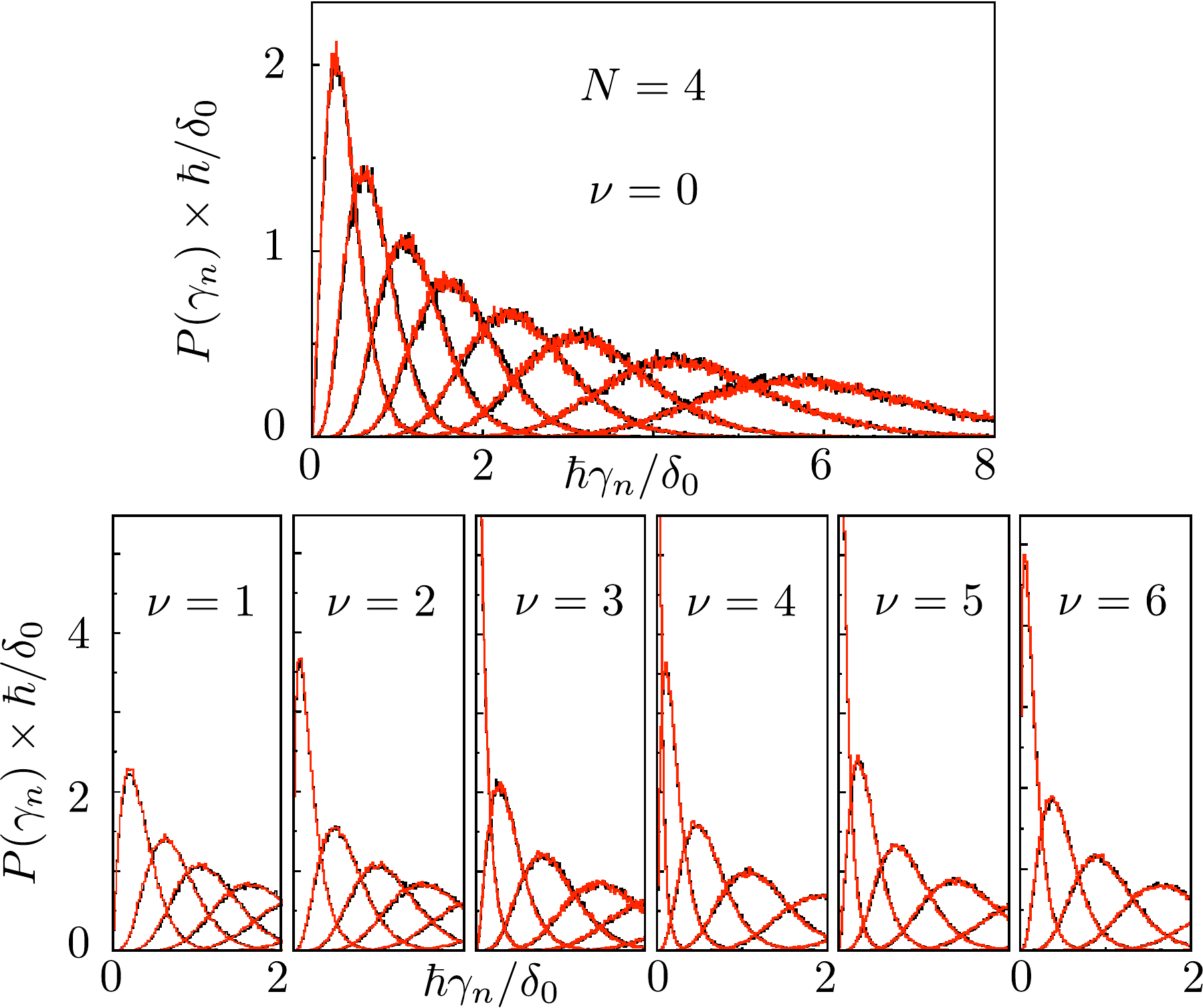}}
\caption{Probability distributions in symmetry class BDI ($\beta=1$) of the $n$-th inverse delay time $\gamma_n$, ordered from small to large: $0<\gamma_1<\gamma_2\cdots<\gamma_{2N}$, with $N=4$. The various plots are for different numbers $\nu=0,1,2,\ldots 6$ of Majorana zero modes. The black histograms of the chiral Gaussian ensemble \eqref{calHdef} (calculated for ${\cal N}=80$) are almost indistinguishable from the the red histograms of the Wishart ensemble, validating our theory. The divergent peak of $P(\gamma_1)$ for $\nu=3,4,5$ is responsible for the divergence of the average density of states \eqref{rhoQ} when the number of zero modes differs by less than two units from the number of channels.
}
\label{fig_gamman}
\end{figure}

To test our analysis, we have numerically generated random matrices from the chiral Gaussian ensemble, on the one hand, and from the Wishart ensemble, on the other hand, and compared the resulting time delay matrices. We find excellent agreement of the delay-time statistics for all three values of the symmetry index $\beta\in\{1,2,4\}$, representative plots for $\beta=1$ are shown in Fig.\ \ref{fig_gamman}.

In view of Eq.\ \eqref{rhoQ} we can directly apply the delay-time distribution to determine the density $\rho(E)$ of quasi-bound states in the Andreev billiard. This is the density of states in the continuous spectrum. For $|\nu|>N$ the full density of states contains additionally a contribution $(|\nu|-N)\delta(E)$ from the discrete spectrum of zero modes that are not coupled to the continuum \cite{note3}.

The probability distribution of the Fermi-level density of states $\rho_0=\rho(0)$ follows upon integration of Eq.\ \eqref{Pgammapmdef}. The ensemble average $\langle\rho_0\rangle$ has a closed-form expression (for details of the calculation see the Appendix),
\begin{equation}
\delta_0\langle \rho_0 \rangle=\begin{cases}
\frac{N(N+1-2/\beta)+\nu^2}{(N+1-2/\beta)^2-\nu^2}, &{\rm if}\;|\nu|<N+1-2/\beta, \\
\frac{N}{|\nu|- N +1-2/\beta}, &{\rm if}\; |\nu| > N -1+2/\beta.
\end{cases}\label{rho0average}
\end{equation}
For $\beta=1$, $|\nu|\in\{N,N\pm 1\}$ and for $\beta=2$, $|\nu|=N$ the average of $\rho_0$ diverges. (There is no divergency for $\beta=4$.) Notice that the $|\nu|-N$ uncoupled zero modes still affect the density of states coupled to the continuum, because they repel the quasi-bound states away from the Fermi level.

\begin{figure}[tb]
\centerline{\includegraphics[width=0.7\linewidth]{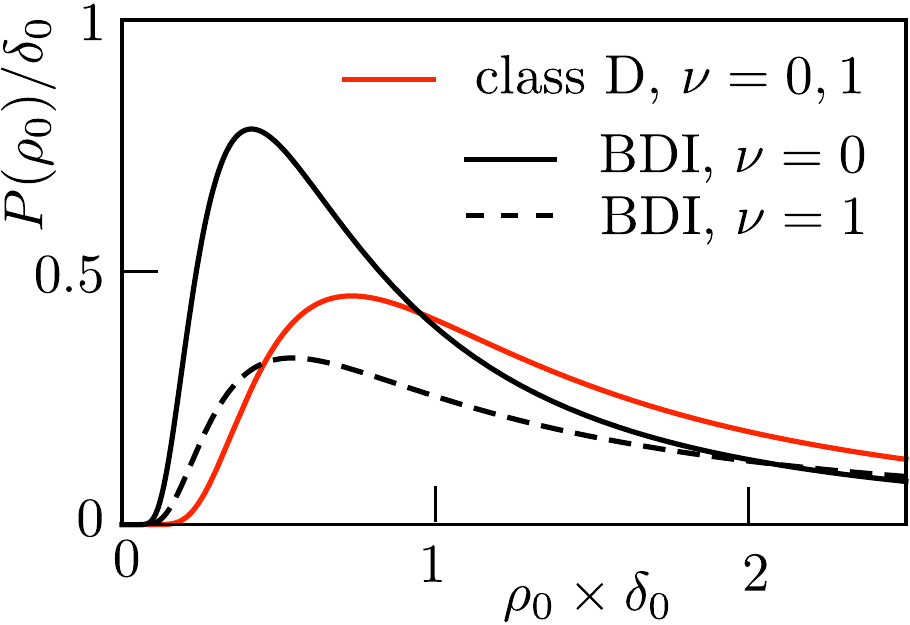}}
\caption{Probability distribution of the Fermi-level density of states, calculated from Eqs.\ \eqref{PDresult} and \eqref{PBDIresult} in symmetry class D (only particle-hole symmetry) and class BDI (particle-hole with chiral symmetry). In class D there is no dependence on the presence or absence of Majorana zero modes \cite{Mar14}, while in class BDI there is.
}
\label{fig_trace}
\end{figure}

As a concrete example we return to the Andreev billiard at the surface of a topological insulator of Fig.\ \ref{fig_layout}, and contrast the delay-time distribution at the Dirac point [chemical potential $\mu=0$ in the Hamiltonian \eqref{HBDG}] and away from the Dirac point ($\mu\gg N\delta_0$). Away from the Dirac point the symmetry class is D (only particle-hole symmetry), while at the Dirac point the additional chiral symmetry promotes the system to class BDI. To simplify the comparison between these two cases we take a point contact with one electron and one hole mode ($N=1$). The scattering matrix has dimension $2\times 2$ and there are two delay times $\tau_1$, $\tau_2$.

The class-D distribution is independent of the presence or absence of Majorana zero modes \cite{Mar14},
\begin{equation}
P_{\rm D}(\tau_1,\tau_2)\propto(\tau_1\tau_2)^{-3}|\tau_1-\tau_2|e^{-(\tau_{\rm H}/2)(1/\tau_1+1/\tau_2)}.\label{PDresult}
\end{equation}
In contrast, the class-BDI distribution \eqref{Pgammapmdef} is sensitive to the number $|\nu|$ of Majorana zero modes,
\begin{align}
&P_{\rm BDI}(\tau_1,\tau_2)\propto e^{-(\tau_{\rm H}/4)(1/\tau_1+1/\tau_2)} \nonumber\\
&\quad\times\begin{cases}
(\tau_1\tau_2)^{-2} & {\rm for}\;\;\nu=0,\\
(\tau_1\tau_2)^{-2-|\nu|/2}|\tau_1-\tau_2| & {\rm for}\;\;|\nu|\geq 1.
\end{cases}\label{PBDIresult}
\end{align}
The corresponding probability distributions of the Fermi-level density of states $\rho_0=\tau_1/\delta_0+\tau_2/\delta_0$ are plotted in Fig. \ref{fig_trace}. Chiral symmetry has a strong effect even for unpaired Majorana zero modes: While away from the Dirac point (class D) the distribution $P(\rho_0)$ is the same for $\nu=0,1$, at the Dirac point (class BDI) these two distributions are significantly different.

In conclusion, this paper presents the solution to a long-standing problem in the theory of chaotic scattering: the effect of chiral symmetry on the statistics of the Wigner-Smith time-delay matrix $Q$. The solution completes a line of investigation in random-matrix theory started six decades ago \cite{Wig67,Dys62}, by establishing the connection between $Q$ and Wishart matrices for the chiral counterparts of the Wigner-Dyson ensembles \cite{Ver93,Ver00}. The solution predicts an effect of Majorana zero modes on the quantum delay-times for chaotic scattering, with significant consequences for the density of states (Fig.\ \ref{fig_trace}). Because the experimental search for Majorana zero modes operates on 1D and 2D systems with chiral symmetry, the general and exact results obtained here are likely to  provide a reliable starting point for more detailed investigations.

We have benefited from discussions with P. W. Brouwer. This research was supported by the Foundation for Fundamental Research on Matter (FOM), the Netherlands Organization for Scientific Research (NWO/OCW), and an ERC Synergy Grant.



\appendix


\section{Details of the calculation of the  Wigner-Smith time-delay distribution in the chiral ensembles}


\subsection{Wishart matrix preliminaries}

Wishart matrices originate from multivariate statistics \cite{Wis28}. We collect some formulas we need \cite{For10}.

The Hermitian positive definite matrix $WW^\dagger$ is called a Wishart matrix if the $n\times m$ ($m\geq n$) rectangular matrix $W$ has real ($\beta=1$), complex ($\beta=2$), or quaternion ($\beta=4$) matrix elements with a Gaussian distribution. For unit covariance matrix, $\langle W_{ij}^{\vphantom{\ast}}W_{i'j'}^\ast\rangle=\delta_{ii'}\delta_{jj'}$, the distribution reads
\begin{equation}
P(W)\propto\exp\left(-\tfrac{1}{2}\beta\,{\rm Tr}\,WW^\dagger\right).
\end{equation}
The eigenvalues of $WW^\dagger$ have the probability distribution
\begin{align}
&P(\lambda_1,\lambda_2,\ldots\lambda_n)\propto\prod_{j>i=1}^n|\lambda_i-\lambda_j|^\beta\nonumber\\
&\quad\times\prod_{k=1}^n\lambda_k^{\beta/2-1}\lambda_k^{\beta(m-n)/2}e^{-\beta\lambda_k/2},\;\;\lambda_k>0.\label{PWishart}
\end{align}
The distribution \eqref{PWishart} is called Wishart distribution, or Laguerre distribution because of its connection with Laguerre polynomials.

\subsection{\label{appb} Degenerate perturbation theory}

We seek the eigenvalue distribution of the $2N\times2N$-dimensional Wigner-Smith time-delay matrix
\begin{equation}
Q_0=-i\hbar S^{\dagger}\frac{dS}{dE}=2\pi \hbar\,\Omega\Omega^\dagger,\;\;
\Omega=K(H+i\pi K^\dagger K)^{-1}.
\label{eq:d}
\end{equation}
As explained in the main text, the key step that allows us to make progress is to invert the order of $\Omega$ and $\Omega^\dagger$, and to consider a larger matrix that is block-diagonal:
\begin{subequations}
\begin{align}\label{eq:q1}
\tilde Q_0&=2\pi\hbar\, \Omega^\dagger\Omega
=2\pi\hbar(\Lambda_-^{-1}\oplus \Lambda_+^{-1}),\\
\Lambda_-=&\pi^2 K_1^\dagger K_1+\mathcal{A}(K_2^\dagger K_2+\epsilon)^{-1}\mathcal{A}^\dagger,
\\
\Lambda_+=&\pi^2 K_2^\dagger K_2+\mathcal{A}^\dagger(K_1^\dagger K_1+\epsilon)^{-1}\mathcal{A}.
\end{align}
\end{subequations}
In this way we can separate the chirality sectors from the very beginning, which is a major simplification.

The two matrices $Q_0$ and $\tilde{Q}_0$ have the same set of nonzero eigenvalues, and $\tilde{Q}_0$ has an additional set of eigenvalues that are identically zero. The corresponding diverging eigenvalues of $\Lambda_\pm$ need to be separated from the finite eigenvalues that determine the inverse delay times $\gamma_n^\pm$. We assume $|\nu|\leq N$ and handle the case $|\nu|> N$ at the end.

To simplify the notation we scale the chiral blocks in the Hamiltonian \eqref{calHdef} as $\mathcal{A}=(2{\cal N}\delta_0/\pi)a$, where $a$ has the Gaussian distribution
\begin{equation}
P(a)\propto\exp\left(-\tfrac{1}{2}\beta{\cal N}\,{\rm Tr}\,aa^\dagger\right).\label{Padef}
\end{equation}
We scale the coupling matrix as $K_i = (2{\cal N}\delta_0/\pi^2)^{1/2}P_i$. The rank-$N$ projector onto the open channels in chirality sector $i=1,2$ is $P_i^{\rm T}P_i^{\vphantom{\rm T}}$, with $P_i^{\vphantom{\rm T}}P_i^{\rm T}=\openone_N$.

To access the finite eigenvalues of $\Lambda_\pm$, we need to perform degenerate perturbation theory in the null spaces of
\begin{equation}
 \Lambda_-^{(0)}=a(\openone_{\mathcal{N}+\nu}-P_2^{\rm T}P_2^{\vphantom{\rm T}})a^\dagger,\;\;
 \Lambda_+^{(0)}=a^\dagger(\openone_{\mathcal{N}}-P_1^{\rm T}P_1^{\vphantom{\rm T}})a,
\end{equation}
with perturbation
\begin{equation}
\begin{split}
&\delta\Lambda_-=2{\cal N}\delta_0(P_1^{\rm T}P_1^{\vphantom{\rm T}}+
a P_2^{\rm T}P_2^{\vphantom{\rm T}} a^\dagger),\\
&\delta\Lambda_+=2{\cal N}\delta_0( P_2^{\rm T}P_2^{\vphantom{\rm T}}+a^\dagger P_1^{\rm T}P_1^{\vphantom{\rm T}} a).
\end{split}
\end{equation}

The null space of $ \Lambda_\pm^{(0)}$ has rank $N_\pm=N\pm \nu\geq 0$. To project onto this null space we make an eigenvalue decomposition,
\begin{equation}
\Lambda_-^{(0)}=u_-^{\vphantom{\dagger}}s_-^{\vphantom{\dagger}}u_-^\dagger,\;\;
\Lambda_+^{(0)}=u_+^{\vphantom{\dagger}}s_+^{\vphantom{\dagger}}u_+^\dagger.
\end{equation}
The matrix $u_\pm$ is unitary and $s_\pm$ is a diagonal matrix with nonnegative entries in descending order. The last $N_\pm=N\pm\nu$ entries on the diagonal of $s_\pm$ vanish, so the projector $p_\pm$ onto the null space consists of the last $N_\pm$ columns of $u_\pm$.  The dimensionalities of $p_+$ and $p_-$ are $({\cal N}+\nu)\times N_+$ and ${\cal N}\times N_-$, respectively. For later use we note that the null space condition $p_\pm^\dagger\Lambda_\pm^{(0)}=0=\Lambda_\pm^{(0)}p_\pm$ requires that
\begin{equation}
 P_2^{\rm T}P_2a^\dagger p_-=a^\dagger p_-,\;\;
P_1^{\rm T}P_1 a p_+=ap_+.\label{Pnullcond}
\end{equation}

The $N_\pm$ finite eigenvalues of $\Lambda_\pm$ are the eigenvalues of the projected perturbation $p_\pm^\dagger\delta\Lambda_\pm^{\vphantom{\dagger}} p_\pm^{\vphantom{\dagger}}$, which we decompose as
\begin{align}
&p^\dagger_\pm  \delta\Lambda_\pm^{\vphantom{\dagger}} p_\pm^{\vphantom{\dagger}}= 2{\cal N}\delta_0
(X_\pm^{\vphantom\dagger}  X_\pm^\dagger + Y_\pm^{\vphantom\dagger}  Y_\pm^\dagger),\label{eq:tilder3}
\\
\begin{split}
&X_-=p_-^\dagger P_1^{\rm T},\;\;X_+=p_+^\dagger P_2^{\rm T},\\
&Y_-=p_-^\dagger aP_2^{\rm T},\;\;Y_+=p_+^\dagger a^\dagger P_1^{\rm T}.
\end{split}
\label{XYpmdef}
\end{align}
The dimensionality of $X_\pm$ and $Y_\pm$ is $N_\pm\times N$. The null space condition \eqref{Pnullcond} implies the constraint
\begin{equation}\label{eq:correl3}
X_-^{\vphantom\dagger} Y_+^\dagger=Y_-^{\vphantom\dagger} X_+^\dagger.
\end{equation}

It is helpful to rescale and combine $X_\pm$, $Y_\pm$ into a single matrix $W_\pm$ of dimension $N_\pm\times 2N$,
\begin{equation}
W_+=\sqrt{\frac{{\cal N}\delta_0}{\pi\hbar}}\,\biggl( X_+,\;Y_+\biggr),\;\;
W_-=\sqrt{\frac{{\cal N}\delta_0}{\pi\hbar}}\,\biggl( -Y_-,\;X_-\biggr).
\end{equation}
The eigenvalues of $W_\pm^{\vphantom{\dagger}} W_\pm^\dagger$ equal the inverse delay times $\gamma_n^\pm$ and the constraint \eqref{eq:correl3} now reads
\begin{equation}
W_-^{\vphantom\dagger}W_+^\dagger=0.\label{Zpmconstraint}
\end{equation}

Considering first the marginal distributions $P_\pm(W_\pm)$ of $W_+$ and $W_-$ separately, we see that these matrices are constructed from rank-$N$ sub-blocks taken from rank-${\cal N}$ random unitary matrices $u_\pm$ and Gaussian matrices $a$. In the limit ${\cal N}\rightarrow\infty$ at fixed $N$ the marginal distributions of $W_\pm$ tend to a Gaussian,
\begin{equation}
P_\pm(W_\pm)\propto\exp\left(-\frac{\beta\pi\hbar}{2\delta_0}\,{\rm Tr}\,W_{\pm}^{\vphantom{\dagger}}W_{\pm}^\dagger\right).
\end{equation}
In view of Eq.\ \eqref{PWishart}, the eigenvalues of $W_\pm^{\vphantom{\dagger}} W_\pm^\dagger$ then have marginal distributions $P_\pm(\{\gamma_n^\pm\})$ of the Wishart form \eqref{Pgammapmdef}.

It remains to show that the two sets of eigenvalues $\gamma_n^+$ and $\gamma_n^-$ have independent distributions, so that
\begin{equation}
P(\{\gamma_n^\pm\})=P_+(\{\gamma_n^+\})P_-(\{\gamma_n^-\}).
\end{equation}
The two matrices $W_+$ and $W_-$ are not independent, because of the constraint \eqref{Zpmconstraint}. To see that this constraint has no effect on the eigenvalue distributions, we make the singular value decomposition
\begin{equation}
W_\pm=\omega_\pm\, \left({\rm{diag}}\left(\sqrt{\gamma_n^\pm}\right),\emptyset_{N_\pm, (2N-N_\pm)}\right)\,\Omega_\pm^\dagger.
\end{equation}
The unitary matrices $\omega_\pm$ and $\Omega_\pm$ have dimension $N_\pm\times N_\pm$ and $2N\times 2N$, respectively, and $\emptyset_{n,m}$ is the $n\times m$ null matrix. The constraint \eqref{Zpmconstraint} is now expressed exclusively in terms of the matrices $\Omega_\pm$ --- the first $N_-$ columns of $\Omega_-$ have to be orthogonal to the first $N_+$ columns of $\Omega_+$. The matrix products
\begin{equation}
W_\pm^{\vphantom{\dagger}} W_\pm^\dagger=\omega_\pm\,{\rm diag}\,(\gamma_n^\pm)\omega_\pm^\dagger
\end{equation}
thus have independent Wishart distributions.

All of this is for $|\nu|\leq N$. The extension to $|\nu|>N$ goes as follows. For $\nu>N$ one has $N_-=0$, so we deal only with a single set of delay times, obtained as the $N_+=2N$ eigenvalues of the Wishart matrix $W_+^\dagger W_+^{\vphantom{\dagger}}$. (We have inverted the order, because $W_+^{\vphantom{\dagger}}W_+^\dagger$ has a spurious set of $\nu$ vanishing eigenvalues, representing zero modes that are uncoupled to the continuum.) Similarly, for $\nu<-N$ one has $N_+=0$ and the delay times are the $N_-=2N$ eigenvalues of the Wishart matrix $W_-^\dagger W_-^{\vphantom{\dagger}}$. The resulting eigenvalue distribution is Eq.\ \eqref{Pgammadefsaturated}.

\subsection{Numerical test}

\begin{figure}[tb]
\centerline{\includegraphics[width=0.7\linewidth]{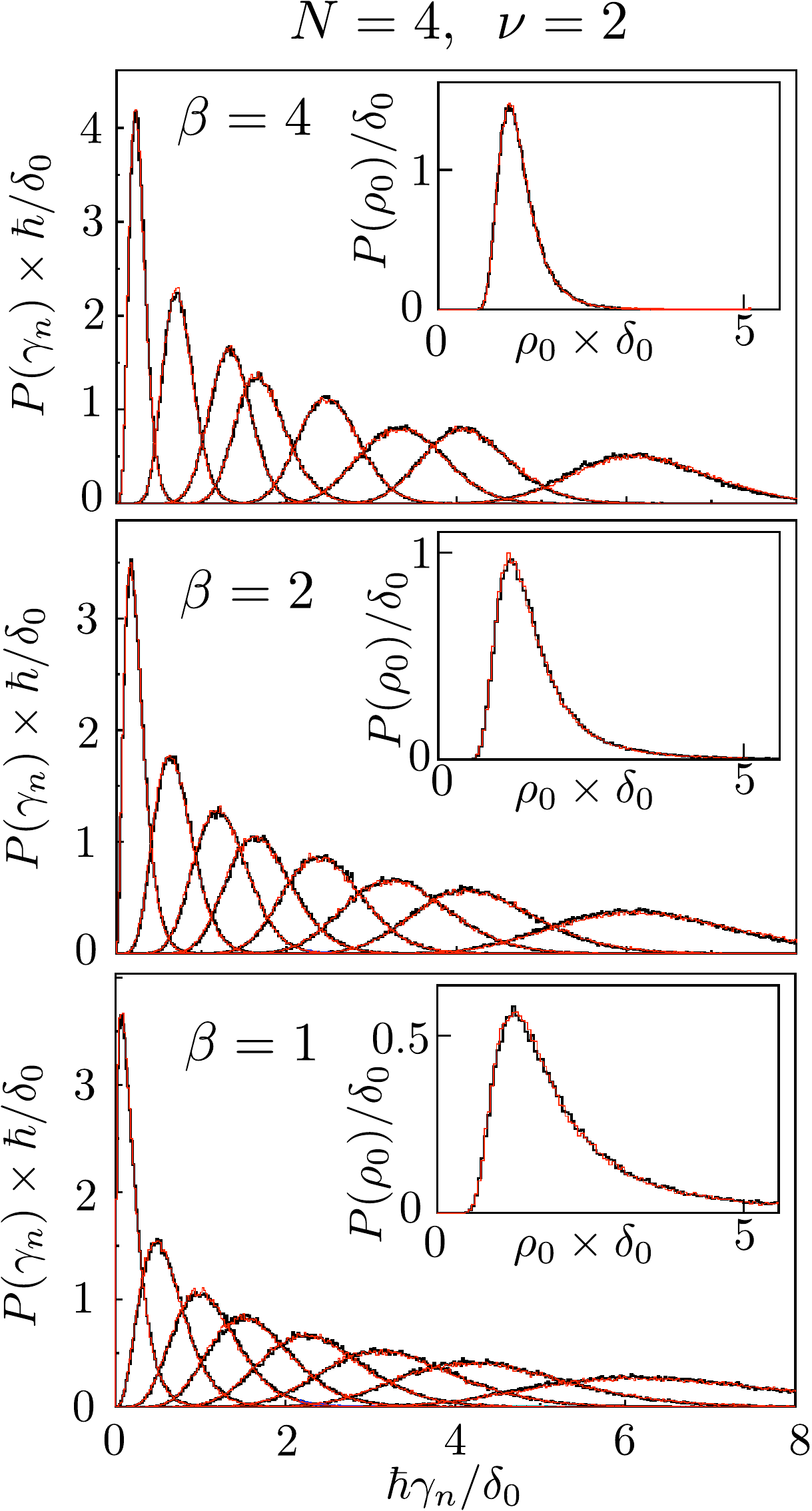}}
\caption{Probability distributions in symmetry class BDI ($\beta=1$), class AIII ($\beta=2$), and class CII ($\beta=4$) of the $n$-th inverse delay time $\gamma_n$, ordered from small to large: $0<\gamma_1<\gamma_2\cdots<\gamma_{2N}$, with $N=4$. All plots are for $\nu=2$ Majorana zero modes. The black histograms of the chiral Gaussian ensemble \eqref{calHdef} (calculated with ${\cal N}=80$ for $\beta=1,2$ and ${\cal N}=120$ for $\beta=4$) are almost indistinguishable from the red histograms of the Wishart ensemble. In each panel the inset shows the corresponding probability distribution of the density of states $\rho_0=\sum_n(2\pi\hbar\gamma_n)^{-1}$.
}
\label{fig_test}
\end{figure}

We have performed extensive numerical simulations to test our analytical result of two independent Wishart distributions for the inverse delay times, comparing with a direct calculation using the Gaussian ensemble of random Hamiltonians. Some results for $\beta=1$, symmetry class BDI are show in the main text (Fig.\ \ref{fig_gamman}), some more results for all three chiral symmetry classes are shown in Fig.\ \ref{fig_test}. The quality of the agreement (the two sets of histograms are almost indistinguishable) convinces us of the validity of our analysis.

\subsection{\label{appc}Generalization to unbalanced coupling}

The results in Appendix \ref{appb} pertain to the case of an equal number $N_1=N_2=N$ of channels coupling to each chiral sector. This is the appropriate case in the context of superconductivity, where the chirality refers to the electron and hole degrees of freedom --- which are balanced under most circumstances. In other contexts, in particular when the chirality refers to a sublattice degree of freedom, the coupling may be unbalanced. We generalize our results to that case.

When $N_1=N_2+\delta N$ Eq.\ \eqref{nudef} for the topological invariant should be replaced by
\begin{equation}
\nu_0=\tfrac{1}{2}\,{\rm Tr}\,S_0={\rm max}\,\left[-\tfrac{1}{2}N_{\rm tot},{\rm min}\bigl(\nu+\tfrac{1}{2}\delta N, \tfrac{1}{2}N_{\rm tot}\bigr)\right].
\end{equation}
The unitary and Hermitian matrix $S_0$ has dimension $2N_{\rm tot}\times 2N_{\rm tot}$, with $N_{\rm tot}=N_1+N_2$. When $N_{\rm tot}$ is odd the number $\nu_0$ is half-integer. The winding number $\nu$ is always an integer.

Because $S_0$ stills commutes with the time-delay matrix $Q_0$ we still have two sets of inverse delay times $\gamma_n^\pm$, associated to the $N_\pm=N_{\rm tot}/2\pm\nu_0$ eigenvalues of $S_0$ equal to $\pm 1$. The two sets again have independent Wishart distributions,

\begin{align}
P_\pm(\{\gamma_n^\pm\})\propto{}&\prod_{j>i=1}^{N_\pm}|\gamma^\pm_i-\gamma^\pm_j|^\beta \prod_{k=1}^{N_\pm} (\gamma_k^\pm)^{\beta/2-1}e^{-\beta\tau_{\rm H}\gamma^\pm_k/4}\nonumber\\
&\qquad\qquad\times (\gamma_k^\pm)^{(\beta/4)|N_{\rm tot}\mp\delta N\mp2\nu|}.\label{eq:plambdagen}
\end{align}
This formula also applies to the saturation regime $|2\nu+\delta N|>N_{\rm tot}$, where either $N_+$ or $N_-$ vanishes and only one set of delay times remains. In this regime the system has an additional $|\nu+\delta N/2|-N_{\rm tot}/2$ zero modes that are not coupled to the continuum.

We can use Eq.\ \eqref{eq:plambdagen} to make contact with the ``single-site limit'' $N_1=1$, $N_2=0$ studied by Fyodorov and Ossipov \cite{Fyo04}. We distinguish positive and negative winding number $\nu$. For $\nu\geq 0$ one has $\nu_0=1/2$, $N_+=1$, $N_-=0$. The single delay time $\tau\equiv 1/\gamma_1^+$ has distribution
\begin{align}
P(\tau)\propto \tau^{-(\beta/2)(1+\nu)-1}e^{-\beta\tau_{\rm H}/4\tau}, \quad\nu\geq 0,
\end{align}
in agreement with Ref.\ \cite{Fyo04} for $\beta=2$. There are then $\nu$ zero modes not coupled to the continuum.

For negative $\nu$ (or equivalently, positive $\nu$ with $N_1=0$, $N_2=1$) Ref.\ \cite{Fyo04} argues that all delay times diverge, but instead we do find one finite $\tau\equiv 1/\gamma_1^-$ with distribution
\begin{align}
P(\tau)\propto \tau^{(\beta/2)\nu-1}e^{-\beta\tau_{\rm H}/4\tau}, \quad\nu\leq -1,
\end{align}
accompanied by $|\nu|-1$ zero modes not coupled to the continuum.

\subsection{Calculation of the average density of states}

The formula \eqref{rho0average} for the ensemble averaged density of states results upon integration of
\begin{align}
&2\pi\hbar\delta_0\langle\rho_0\rangle=\int_0^\infty d\gamma^+_1\cdots\int_0^\infty d\gamma^+_{N_+}P_+(\{\gamma_n^+\})\sum_{n=1}^{N_+}\frac{1}{\gamma_n^+} \nonumber\\
&\quad+\int_0^\infty d\gamma^-_1\cdots\int_0^\infty d\gamma^-_{N_-}P_-(\{\gamma_n^-\})\sum_{n=1}^{N_-}\frac{1}{\gamma_n^-},
\end{align}
with probability distributions $P_\pm$ given by Eq.\ \eqref{Pgammapmdef}. These integrals can be carried out in closed form, as follows.

We need to evaluate an expression of the form
\begin{equation}
I=\frac{1}{C}\prod_{k=1}^N\int_0^\infty d\gamma_k\, \gamma_k^p e^{-\beta\tau_{\rm H}\gamma_k/4} \prod_{j>i=1}^{N}|\gamma_i-\gamma_j|^\beta \left(\sum_{n=1}^N\frac{1}{\gamma_n}\right),
\end{equation}
with normalization integral
\begin{equation}
C=\prod_{k=1}^N\int_0^\infty d\gamma_k\, \gamma_k^p e^{-\beta\tau_{\rm H}\gamma_k/4} \prod_{j>i=1}^{N}|\gamma_i-\gamma_j|^\beta.
\end{equation}
For a finite answer we need an exponent $p>0$.

We substitute $\gamma_k^{p-1}=p^{-1}d\gamma_k^p/d\gamma_k$ and perform a partial integration,
\begin{align}
I={}&\frac{1}{pC}\prod_{k=1}^N\int_0^\infty d\gamma_k\, e^{-\beta\tau_{\rm H}\gamma_k/4} \prod_{j>i=1}^{N}|\gamma_i-\gamma_j|^\beta \nonumber\\
&\times\left(\sum_{n=1}^N\frac{d}{d\gamma_n}\right)\prod_{k'=1}^N\gamma_{k'}^p\\
={}&\frac{\beta N\tau_{\rm H}}{4p}-\frac{1}{pC}\prod_{k=1}^N\int_0^\infty d\gamma_k\, \gamma_k^p e^{-\beta\tau_{\rm H}\gamma_k/4}\nonumber\\
&\times \left(\sum_{n=1}^N\frac{d}{d\gamma_n}\right)\prod_{j>i=1}^{N}|\gamma_i-\gamma_j|^\beta\nonumber\\
={}&\frac{\beta N\tau_{\rm H}}{4p},
\end{align}
because
\begin{equation}
\left(\sum_{n=1}^N\frac{d}{d\gamma_n}\right)\prod_{j>i=1}^{N}|\gamma_i-\gamma_j|^\beta=0.
\end{equation}


\begin{thebibliography}{24}
\bibitem{note1} The distribution \eqref{Pgammadef} is known as a Laguerre distribution in random-matrix theory. It represents the eigenvalue distribution of a Wishart matrix $WW^\dagger$ for $\beta=1$ (when $W$ is a real Gaussian $N\times(2N+1)$-dimensional matrix) and for $\beta=2$ (complex Gaussian $N\times 2N$ matrix $W$). For $\beta=4$ there is no corresponding Wishart ensemble.
\bibitem{note0} According to the ``ten-fold way'' classification of topological states of matter \cite{Has10,Qi11,Ryu10,Alt97}, class BDI is nontrivial in 1D but not in 2D. To reconcile this with the 2D realization of Fig.\ \ref{fig_layout}, we refer to the analysis of Teo and Kane \cite{Teo10}, who showed that the effective dimensionality for a topological defect is $d-d'$, where $d=2$, $d'=1$ for a vortex on the surface of a topological insulator. More generally, $d$ is the dimensionality of the Brillouin zone and $d'$ is the dimensionality of a contour that encloses the defect.
\bibitem{spincounting} The number $2N$ of scattering channels includes a factor of $2$ from the electron-hole degree of freedom. For $\beta=4$ each scattering channel (and hence each delay time $\tau_n$) has a twofold Kramers degeneracy from the spin degree of freedom, while for $\beta=1,2$ the spin degree of freedom is counted separately in $N$. The mean level spacing $\delta_0$ refers to distinct levels in the bulk of the spectrum (away from $E=0$), including electron-hole and spin degrees of freedom but not counting degeneracies.
\bibitem{note5} The Laguerre distributions \eqref{Pgammapmdef} and \eqref{Pgammadefsaturated} are the eigenvalue distributions of a Wishart matrix $WW^\dagger$ when $W$ has dimension $(N\pm\nu)\times 2N$ for $|\nu|<N$ and dimension $2N\times(N+|\nu|)$ for $|\nu|\geq N$.
\bibitem{note3} The $|\nu|-N$ uncoupled zero modes in the Andreev billiard, not broadened by the $2N$ scattering channels into the continuum, span the null-space of ${\cal H}+i\pi K^\dagger K$. For $|\nu|\leq N$ all zero modes are broadened by coupling to the continuum.
\end{thebibliography}
\end{document}